# Dynamics of a Dust Crystal with Two Different Size Dust Species

L.S. Matthews, K. Qiao, T.W. Hyde

*CASPER (Center for Astrophysics, Space Physics, and Engineering Research), Baylor University, Waco, Texas 76798-7310, USA (Truell_Hyde@baylor.edu)*

**Abstract**

A self-consistent three-dimensional model for a complex (dusty) plasma is used to study the effects of multiple-sized dust grains in a dust crystal. In addition to the interparticle forces, which interact through a Yukawa potential, the model includes the effects of gravity, the variation of the sheath potential above the powered electrode, and a radial confining potential. Simulations studied various ratios of a mix of 6.5- and 8.9-micron monodisperse particles and compared their correlation functions, electric potential energy of the crystal formations, and the dispersion relations for in-plane and out-of-plane dust lattice wave (DLW) modes for two different sheath thicknesses. In the 7 mm sheath, the particles formed two layers in the vertical direction by size, and acted as a two-layer crystal with weak correlation between the layers. In the 3 mm sheath, the particles formed an essentially monolayer crystal; however the crystal dynamics showed some characteristics of a bilayer crystal.

## 1. Background

The two-dimensional (2D) plasma crystal produced in a typical laboratory complex plasma environment is formed when negatively charged dust particles are levitated in the sheath region of an rf discharge plasma (Chu and I, 1994; Hayashi and Tachibana, 1994; Thomas et al., 1994). The system of particles is usually constrained in the horizontal direction by the force of an inwardly directed electric field. The negative charge on a particle is shielded by the ambient plasma, and it can be assumed that the particles interact with each other through a repulsive Yukawa potential defined by

$$\Phi(r) = q \exp(-r/\lambda_D)/4\pi\varepsilon_o r, \quad (1)$$

where $q$ is the particle charge, $r$ is the distance between any two particles and $\lambda_D$ is the dust Debye length. Given appropriate plasma conditions, namely that the coupling constant $\Gamma$, the ratio of the particles' electric potential energy to their kinetic energy, is greater than some critical value, the dust particles form a hexagonal lattice. In the laboratory the dust crystals are usually two-dimensional or 2 ½-dimensional due to the influence of gravity.

Many of the fundamental properties of plasma crystals, including the particle charge, the Debye length, and the sheath potential, can be obtained from the dispersion relations of the dust lattice wave (DLW) modes (Homann et al., 1998; Nunomura et al., 2002, Qiao and Hyde, 2003). The dispersion relation of a wave mode is defined as the relationship between the angular frequency (or frequency) and the wave number (or wavelength). It provides the fundamental properties of a given wave mode, for example the phase velocity,

$$v_{ph} = \frac{\omega}{k} \quad (2)$$

or the group velocity,

$$v_g = \frac{d\omega}{dk}. \quad (3)$$

The dispersion relation draws its name from the fact that the wave is dispersive since phase velocities are different for different frequencies; in other words, the wave number (or wavelength) depends on the frequency.

To date, studies of crystal dynamics have focused mainly on collections of monodisperse particles, which are found only in artificial environments, such as the laboratory. It would be of great benefit to understand the dynamics of dust collections with polydisperse size distributions, since this would allow the theory to be extended to more realistic environments such as those in which dusty plasmas are generally found. This paper will compare the crystal formation and system dynamics of dust crystals composed of monodisperse 6.5 µm and 8.9 µm particles to that of dust crystals composed of three different ratios of the two sizes. The results are then compared to similar crystals created in the lab.

## 2. Numerical model

The numerical model employed in this study is based on the box_tree code which was developed to model the dynamics of a large number of particles interacting through gravitational and electrostatic forces (Richardson, 1993; Matthews and Hyde, 2003, 2004). The box_tree code is a hybrid of two computer algorithms, a box code and a tree code. The tree code provides a method for a fast calculation of the interparticle forces by means of a multipole expansion. In the z-direction, earth's gravity is balanced by the dc self-bias electric field of the powered electrode in the reference cell, allowing charged particles to levitate in the plasma sheath above the electrode. Studies have shown that the electric potential in the plasma sheath varies with the distance above the electrode and may be approximated by a parabolic potential (Tomme et al., 2000). The sheath thickness and the levitation height of the particles, which in the lab are a function of the power delivered to the cell, can also be adjusted by the steepness of this potential. In addition, a radial parabolic confining potential is used to model the boundary conditions produced by a small lip on the edge of the lower electrode in the cell (Konopka et al., 2000).

## 3. Simulations

Five different cases were examined comparing the effects of bidisperse size distributions on crystal formation and system dynamics to that of monodisperse size distributions. The particles were modeled as $6.50 \pm .08$ μm and $8.89 \pm .09$ μm spheres with a density of 1.51 gm/cm$^3$, which corresponds to the physical characteristics of commercially available melamine formaldehyde spheres used in lab experiments. Models I (a) and (b) correspond to monodisperse collections of 6.5 μm and 8.9 μm spheres, respectively, while Models IIa-IIc correspond to 75/25, 50/50, and 25/75 ratios of 6.5 μm and 8.9 μm spheres. These ratios were chosen to match those used in experiments in our cell (Smith et al., 2005). The charges on particles in laboratory dust crystals under these conditions have been inferred to lie within the range of 10,000e$^-$ to 15,000e$^-$; accordingly, the charge on an 8.90 μm sphere was set to 12,000e$^-$ which corresponds to a surface potential $\Phi_s$ = -3.9 V. Assuming that the only charging currents are the primary electron and ion currents, $\Phi_s$ is independent of particle size and this potential can be used to determine the charge on each particle employing the relation q = $4\pi\varepsilon_0 a \Phi_s e^{-(a/\lambda_D)}$, where $a$ is the radius of the grain. Assuming a plasma density of $10^9$ /cm$^{-3}$ and electron energies of a few eV, which are typical values found in our cell, $\lambda_D$ is usually on the order of $10^2$ μm. In these simulations it was set at 600 μm, corresponding to an electron energy of 9 eV.

For each model, simulations were run for a 3 mm and 7 mm sheath, the approximate sheath thicknesses in our cell for investigating these conditions. For the 7 mm sheath, ~9000 particles were used while ~5000 were modeled in the 3 mm sheath.

The particles were originally randomly placed in a monolayer within a cloud of diameter 3 cm and the simulations then allowed to progress until approximately 5 s of data on the particles in their equilibrium states were collected. The resulting crystal formations were then analyzed by employing voronoi diagrams to determine the crystalline structure and by calculating the normalized potential energy of the crystal,

$$E_N = \frac{1}{4\pi\varepsilon_0 q_{avg}} \sum_{i>j} \frac{q_i q_j}{r_{ij}} e^{(-r_{ij}/\lambda_D)} \quad (4)$$

where $r_{ij}$ is the distance between the i$^{th}$ and j$^{th}$ particle, and $q_{avg}$ is the average particle charge in the dust crystal. The pair correlation function, g(r), which represents the probability of finding two particles separated by a distance $r$, and the dispersion relations for longitudinal, in-plane and out-of-plane transverse DLWs were also calculated for each model.

## 4. Results

*4.1 Equilibrium positions and correlation functions*

In both the 7 mm and the 3 mm sheaths, the two different sizes of particles separated into distinct layers. In the 7 mm sheath, the vertical spacing between layers was approximately 500 μm which is on the same order as the average interparticle spacing (~450 μm). The ensemble behaves as a partially correlated two-layer crystal. In the 3 mm sheath, the vertical separation was approximately 100 µm. Since this distance is smaller than the average interparticle spacing (~500 µm), the two layers interact in a correlated manner and behave almost as a single-layer crystal.

Table 1. Distribution of cells in voronoi diagrams for equilibrium positions of crystals in the 7 mm sheath. Percentages are given for the layers of 6.5 µm and 8.9 µm spheres separately.

|  | 6.5 µm | | | 8.9 µm | | |
| --- | --- | --- | --- | --- | --- | --- |
|  | % 5-sided | % 6-sided | % 7-sided | % 5-sided | % 6-sided | % 7-sided |
| Model Ia | 10.8 | 78.5 | 10.5 | 0 | 0 | 0 |
| Model Ib | 0 | 0 | 0 | 14.5 | 71.4 | 14.0 |
| Model IIa | 8.6 | 83.1 | 8.3 | 12.6 | 75.5 | 11.8 |
| Model IIb | 8.4 | 84.0 | 7.6 | 6.7 | 86.7 | 6.7 |
| Model IIc | 15.6 | 69.8 | 14.7 | 5.8 | 88.5 | 5.6 |

Table 2. Distribution of cells in voronoi diagrams for equilibrium positions of crystals in the 3 mm sheath.

|  | % 5-sided | % 6-sided | % 7-sided |
| --- | --- | --- | --- |
| Model Ia | 4.7 | 91.0 | 4.3 |
| Model Ib | 4.2 | 92.1 | 3.7 |
| Model IIa | 9.5 | 81.5 | 9.0 |
| Model IIb | 13.1 | 74.2 | 12.7 |
| Model IIc | 14.3 | 71.7 | 14.0 |

This difference in behavior can be seen by comparing the number of five-, six-, and seven-sided cells in the voronoi diagrams of the particles' equilibrium positions (Tables 1 and 2). In the 7 mm sheath, each layer was well ordered, with a high percentage of six-sided cells. Because the interparticle spacing for the 6.5 µm and 8.9 µm particles was almost the same for Model IIb, the particles were nearly aligned in the vertical direction (Fig. 1a), even though ion drag was not considered. This resulted in the most stable crystalline formation with the lowest normalized potential energy (Figure 2a).

In contrast, in the 3 mm sheath, the particles were in a single layer and Model IIa had the most stable crystalline structure, as seen in the comparison of the number of six-sided cells in the voronoi diagram (Table 2) and the normalized potential energy of the crystal (Fig 2b). The explanation for this lies in the relative positions of the 8.9 µm and 6.5 µm particles in the different models. In Model IIa, the minority population (8.9 µm particles) were almost evenly interspersed among the 6.5 µm particles (Fig 1b), while in Model IIc the minority population (6.5 µm particles) formed "strings" within the overall crystalline structure (Fig 1c).

This structure described above is also seen in the correlation functions for the 7 mm sheath and the 3 mm sheath (Figs. 3 and 4, respectively). In the 7 mm sheath, the peaks for the minority population occur at larger particle spacing than they do for the majority population in Models IIa and IIc (Figs. 3a and c) and are weakly correlated. The peaks for the two particle sizes in Model IIb appear at approximately the same particle spacing (Fig. 3b). In the 3 mm sheath, for Model IIa the peak with the greatest magnitude for the 8.9 µm particles is aligned with the second peak for the overall crystal (Fig. 4a), which can be explained by the fact that most of the 8.9 μm particles' nearest neighbors are 6.5 μm particles and the nearest 8.9 μm particle is a second-nearest neighbor. In contrast, in Model IIc, the first two peaks for the 6.5 μm particles are roughly of the same magnitude (Fig 4c), since the 6.5 μm particles usually have one to two nearest neighbors that are also 6.5 μm particles. It is interesting to note that in laboratory experiments with a 3 mm sheath, while the crystal structure was not as well-ordered in the bidisperse distribution (a 50/50 mix corresponding to Model IIb), it appeared to be more stable against changes due to varying the power delivered to the cell when compared to a monodisperse collection of dust (Smith et al., 2005).

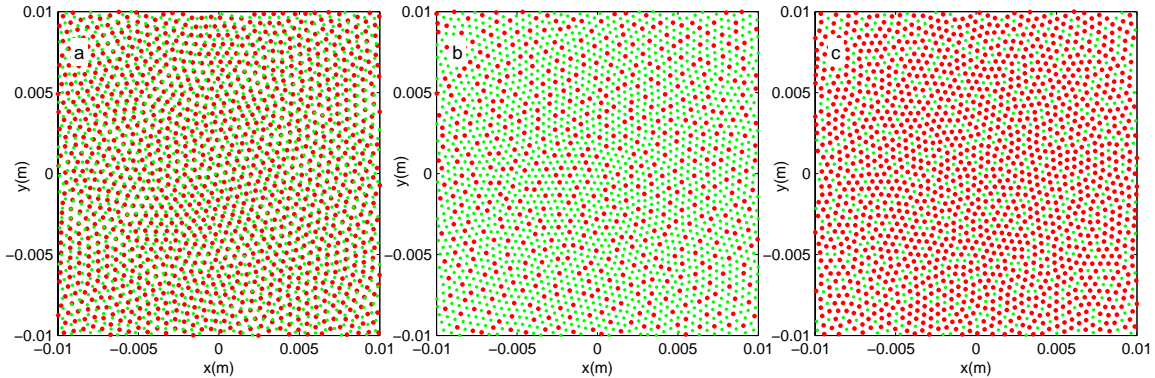

Figure 1. Top view of center of crystal. (a) In the 7mm sheath, Model IIb shows alignment of the 6.5 µm particles (green) with the 9µm particles (red) below them. In the 3 mm sheath, (b) regular arrangement of 8.9 µm particles in Model IIa, (c) "strings" of 6.5 µm particles in Model IIc.

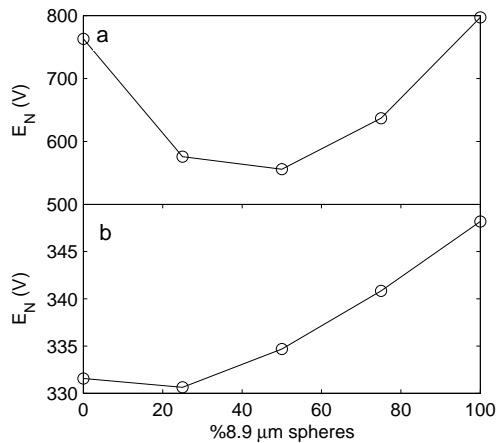

Figure 2. Normalized potential energy of crystal in (a) 7mm sheath and (b) 3mm sheath given as a function of the percent of the population which consists of 8.9 μm spheres.

*4.2. Dispersion Relations*

The dispersion relations for the longitudinal, in-plane transverse and out-of-plane transverse dust lattice waves (DLWs) for the 7 mm sheath are shown in Figures 5-7. The patterns of the intensity graphs show the dispersion relations obtained from simulation, with the solid lines showing the dispersion relations derived from an analytical method assuming a 2D hexagonal lattice and Yukawa potential between dust grains (Qiao and Hyde, 2003). The values of particle charge, mass, radius, the Debye length, and the horizontal and vertical confining potential used in the analytical method are all the same as those used in the simulation.

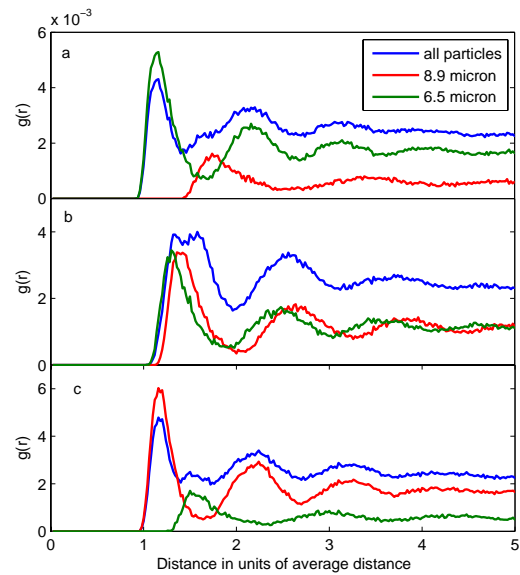

Figure 3. Pair correlation functions for 7 mm sheath, Models IIa-IIc, respectively.

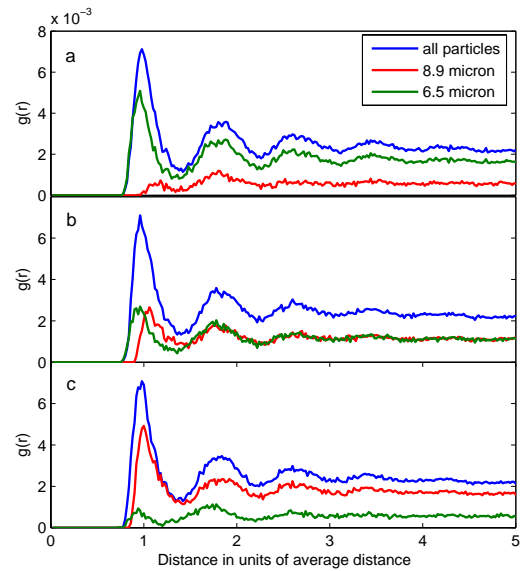

Figure 4. Pair correlation functions for 3 mm sheath, Models IIa-IIc, respectively.

Figure 5 shows the dispersion relations for Model Ia, which only has one layer. The simulation and analytical results agree with each other well for the longitudinal and out-of-plane transverse modes (Fig. 5b, c). For in-plane transverse modes (Fig. 5a), the simulation shows an optical dispersion relation for long wavelengths, which is not predicted by the analytical method based on the assumption of a 2D system. The dispersion relations for Model Ib are similar and are not shown.

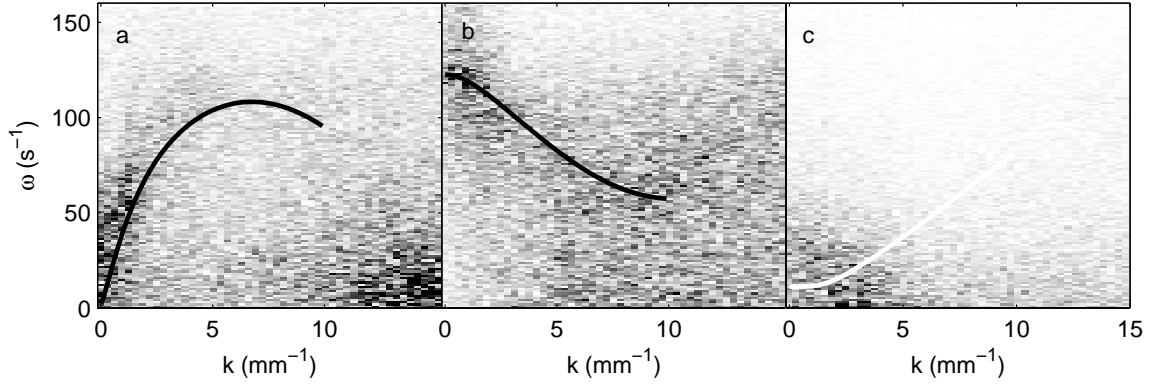

Figure 5. Dispersion relations in the 7mm sheath, Model Ia, (a) in-plane transverse, (b) longitudinal, and (c) out-of-plane transverse waves.

The dispersion relations for the longitudinal and out-of-plane transverse DLWs within individual layers for Model IIa are shown in Figure 6. (The dispersion relations for Models IIb and IIc are similar and are not shown). These simulation results agree with the analytical results for all cases. The difference between the wave properties in the two layers can be seen clearly in the out-of-plane transverse DLW dispersion relations. The dispersion relations are of different values of $\omega$ because the particle mass $m$ and charge $q$ are different for the two layers (Qiao and Hyde, 2003). For Models IIa the dispersion relations for the two layers also have different slopes. This is because of the different interparticle spacing $a$ for the two layers, yielding differing values of the shielding parameter $\kappa$ (Qiao and Hyde, 2003). As an example, the out-of-plane DLW dispersion relation for the two-layer system is shown for Model IIb (Fig. 7). The double dispersion corresponds to the two layers. Analytical results for the dispersion relations for the two layers are superimposed and show good agreement with simulation. Similar results were obtained for Models IIa and IIc.

In the 3 mm sheath, the analytical results also agreed well with the results for the numerical models. The double dispersion for the out-of-plane transverse waves for Models IIa-IIc are shown in Figure 8. It is interesting to note that while the collection of particles appeared to behave as a single-layer crystal, in each case a faint secondary dispersion relation can be seen offset from the dispersion relation for the majority species, or main crystal, suggesting that the crystal has at least some of the characteristics of a two-layer system.

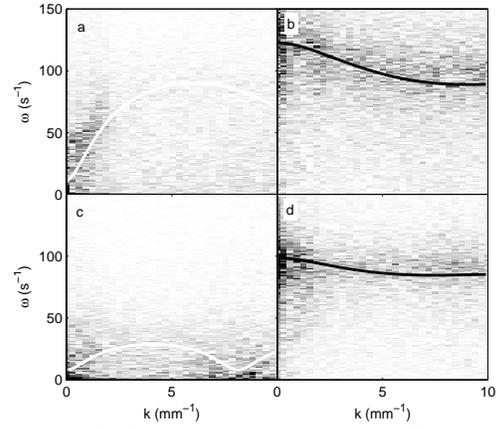

Figure 6. Dispersion relations in the 7mm sheath, Model IIa, longitudinal and out-of-plane transverse waves for top (a, b) and bottom layers (c, d). Note the difference between the two layers in the slopes for the dispersion relations.

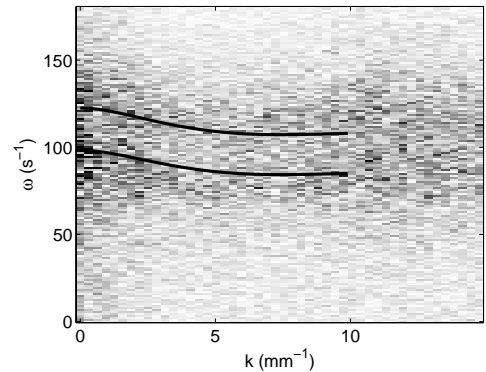

Figure 7. Double dispersion relation seen for the out-of-plane transverse wave for Model IIb in the 7 mm sheath.

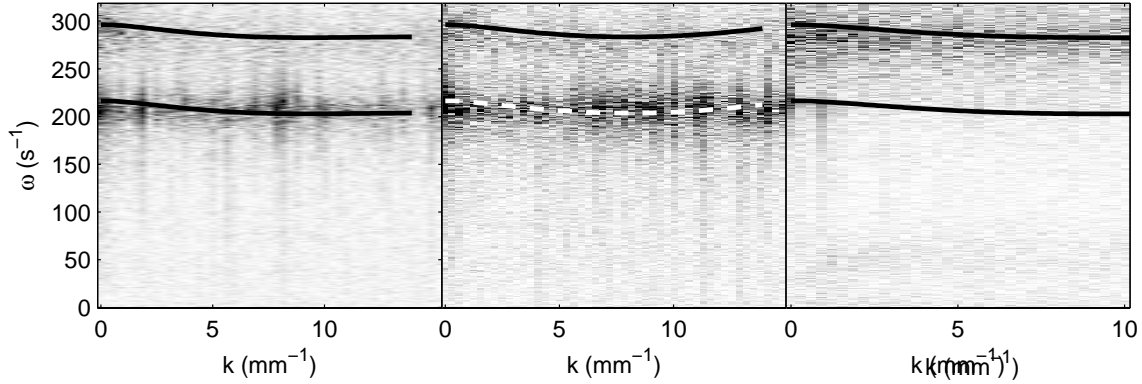

Figure 8. Dispersion relations for out-of-plane transverse waves for (a) Model IIa, (b) Model IIb, and (c) Model IIc in the 3 mm sheath.

## 5. Conclusions

Numerical simulations were run for complex plasma systems containing two different dust size species. The characteristics of the crystal were shown to not only depend on the ratio of the two different dust sizes, but also on the size of the sheath in which the dust was suspended. The larger sheath thickness allowed the particles to separate into vertical layers based on size. Each layer formed a crystalline lattice which only weakly interacted with the other layer. The most stable crystal formation was found for an equal number of 6.5 and 8.9 μm particles, in which case the particles were nearly aligned in the vertical direction. In the smaller sheath, the two different size populations were separated by a vertical distance smaller than their average interparticle spacing. In this case, the particles formed a single monolayer crystal, but the crystal dynamics still showed evidence of weak two-layer characteristics.

Analysis of the potential energy of the crystal shows that in a two layer system, as seen in the 7 mm sheath, a bidisperse collection of particles has a lower energy, and thus is more stable than a collection of monodisperse particles. This is to be expected since the average interparticle spacing for this case is greater in a two-layer system. However, even in the thinner sheath where the resulting structure could be considered a single-layer crystal, the energy of a monodisperse system was greater than that of a dust population with a minority of the population replaced by different-sized particles. Further laboratory investigation of a range of mixtures is needed to investigate this effect more fully.

The dispersion relations for the longitudinal, in-plane and out-of-plane transverse dust lattice waves (DLWs) were obtained for both 3mm and 7mm sheaths. The simulation results for the longitudinal and out-of-plane transverse modes agree well with the results from an analytical model employing the same values of particle charge, mass, radius, Debye length, and horizontal and vertical confining potential, showing that the fundamental properties of plasma crystals can be obtained from the dispersion relations. For in-plane transverse modes, the simulation shows an optical dispersion relation for long wavelengths, which is not predicted by the analytical method based on the assumption of a 2D system.

Double dispersion relations for the out-of-plane transverse DLWs were obtained for simulations of a system consisting of both 6.5 and 8.9 μm particles and showing a two layer structure. The dispersion relations exhibit different values of ω since the particle mass $m$ and charge $q$ are different for the two layers. For simulations of systems with different number of 6.5 and 8.9 μm particles, the two dispersion relations also show different slopes because of the different interparticle spacing $a$ for the two layers. This indicates that a careful experimental examination of the dispersion relations for such a system might well provide a new diagnostic method for distinguishing between layers of different particle sizes for multidisperse collections of particles.

Further studies will investigate the influence of broader size distributions on the formation and resulting dynamics of dust crystals as well as their influence on the sheath potential.